\documentclass[twocolumn]{aastex62}
\usepackage[utf8]{inputenc}

\begin{document}

\title{What Does a Successful Postdoctoral Fellowship Publication Record Look Like?}

\author{Joshua Pepper}
\affiliation{Department of Physics, Lehigh University, 16 Memorial Drive East, Bethlehem, PA, 18015}

\author{Oliwia D. Krupinska}
\affiliation{Department of Physics, Lehigh University, 16 Memorial Drive East, Bethlehem, PA, 18015}

\author{Keivan G. Stassun}
\affiliation{Vanderbilt University, Department of Physics \& Astronomy, 6301 Stevenson Center Ln., Nashville, TN 37235, USA}
\affiliation{Vanderbilt Initiative in Data-intensive Astrophysics (VIDA), 6301 Stevenson Center Lane, Nashville, TN 37235, U}
\affiliation{Fisk University, USA}

\author{Dawn M. Gelino}
\affiliation{NASA Exoplanet Science Institute, Caltech, MS 100-22, 770 South Wilson Avenue, Pasadena, CA 91125, USA}
\affiliation{Former Sagan Fellowship Program Lead, Current NASA Hubble Fellow Program (NHFP) Co-Lead}

\begin{abstract}
Obtaining a prize postdoctoral fellowship in astronomy and astrophysics involves a number of factors, many of which cannot be quantified.  One criterion that can be measured is the publication record of an applicant. The publication records of past fellowship recipients may, therefore, provide some quantitative guidance for future prospective applicants. We investigated the publication patterns of recipients of the NASA prize postdoctoral fellowships in the Hubble, Einstein, and Sagan programs from 2014 through 2017, using the NASA ADS reference system.  We tabulated their publications at the point where fellowship applications were submitted, and we find that the 133 fellowship recipients in that time frame had a median of $6\pm2$ first-author publications, and $14\pm6$ co-authored publications. The full range of first author papers is 1 to 15, and for all papers ranges from 2 to 76, indicating very diverse publication patterns. Thus, while fellowship recipients generally have strong publication records, the distribution of both first-author and co-authored papers is quite broad; there is no apparent threshold of publications necessary to obtain these fellowships.  We also examined the post-PhD publication rates for each of the three fellowship programs, between male and female recipients, across the four years of the analysis and find no consistent trends.  We hope that these findings will prove a useful reference to future junior scientists.
\end{abstract}

\section{Introduction}

As astrophysics students move through their graduate careers en route to a PhD, many will begin preparing to apply to postdoctoral positions. The preparation process poses an inevitable question, “What does an applicant to a postdoctoral position need to accomplish in graduate school in order to be competitive in the job market?” The full portfolio of academic accomplishments is generally reflected in a CV, which typically includes graduate GPA, teaching or TA experience, mentoring or advising of students, conference presentations and proceedings, invited talks and colloquia, roles or leadership positions within collaborations, community and committee service, observing or funding proposals, and publications. In some cases, it may also include graduate courses and/or listings of skills. Of the listed accomplishments, those that form the core of the CV, and are often believed to be the most important achievements for a young scientist, are the peer-reviewed publications. Students may recast the original question and ask, “How many papers do I need to publish in order to get a job after my PhD?” 

There is no single answer to that question.  Different subfields of astronomy may have different expectations for the publications of graduating PhD students. There is no consensus of the relative importance of first authorship versus co-authorship on papers, nor the relative value of later authorship position; is the second author position of greater importance than a later position? There is also no agreement about the use of normalized publication numbers, in which the authorship value decreases with additional co-authors on the paper. Indeed, a publication record is clearly not a complete testament to the capabilities, accomplishments, and research potential of a scientist, but it may serve as a first glance into how the student compares to others within similar research areas. 

It is understood that the sheer number of publications is a crude statistic and does not capture the quality of the research. For that, citation rates are often examined. However, those numbers can also be interpreted in various ways. Should citation rates be normalized? Should self-citations be excluded? Beyond the lack of agreement of the mechanics of such bibliometrics \citep{Wildgaard:2014}, there is also evidence that various biases influence the peer-review process, leading to lower citation rates and numbers of prestigious publications for women \citep{Bendels:2018,Caplar:2017} and leading to an inconsistent comparison between genders. For more about the use and misuse of citation statistics, see \citet{Waltman:2016}, \citet{Kurtz:2017}, and \citet{Hazoglu:2017}. For a discussion of alternate metrics, see \citet{Henneken:2017}.

With all those issues to consider, probably the best advice one can give graduate students is that more publications are better than fewer (although, see \citet{Waltman:2013}), and first-author papers are better than co-authorships. Still, for a student trying to plan their research program in graduate school, it is natural to ask how many papers they need to publish.  Expectations change over time, so it is worth looking at recent trends to see how a successful publication record has looked empirically in recent years.

Although there are obviously many ways to measure professional success in astronomy, one way we can do so is to look at the recipients of prestigious postdoctoral fellowships. These positions are generally awarded to early career astronomers right out of graduate school or within a few years of receiving their PhD.  While some fellowships are independent or tied to certain universities, many prestigious fellowships are awarded by NASA and the NSF. The NSF offers postdoctoral fellowships in astronomy, and NASA has the NASA Postdoctoral Program (NPP), and has also offered prize postdoctoral fellowships over the years under the various categories of the Hubble, Chandra/Fermi/Einstein, and Michelson/Sagan programs.  While the details of those programs have changed somewhat over time, the Hubble fellowship covers stellar and galactic science, the Einstein fellowship focuses on high–energy astronomy and the astrophysics of extreme objects, and the Sagan fellowship focuses on exoplanets and instrumentation.  Currently, the fellowships are aligned with the three primary science categories defined by NASA, with Hubble addressing Cosmic Origins, Einstein connected to Physics of the Cosmos, and Sagan related to Exoplanet Exploration.

These fellowships are highly desirable, and the past recipients of these fellowships may serve as examples of successful scientists at the postdoctoral stage of their career.\footnote{It is important to note that many well-qualified applicants do not receive these fellowships, simply because there is a limited number of available fellowship positions.}  It should be noted that the recipients’ publications are not the only, or even primary materials reviewed by the committees that award these fellowships.  The scientific proposals are likely the most important components of the applications, and letters of recommendation play a significant role as well.  However, for the purposes of defining a population at the post-graduate stage of a scientific career, the recipients of these awards can serve as a reference point which may be used by current graduate students looking for an understanding of the possible range of publication records. 

We have assembled and analyzed the publication records of several years of recipients of the NASA prize fellowships, examining the publication records of the recipients at the time of application to the fellowship, and also at the end of the year they received a PhD. We do not intend to define a specific publication goal for students but rather to describe what the records of typical recipients of these fellowships look like, under the assumption that providing this information is better than ignoring it. We discuss the methods that we used to assemble and describe the population of fellowship recipients in section 2. In section 3 we explain how we assemble publication information for those people. In section 4 we analyze our findings and in section 5 we discuss the implications and limitation of this analysis.

\section{Recipients and Biographical Information}

We used the set of prize NASA postdoctoral fellowships to conduct our analysis. We identified the recipients of the fellowships from years 2014 to 2017, based on the listings from the Hubble\footnote{http://www.stsci.edu/stsci-research/fellowships/nasa-hubble-fellowship-program/past-hubble-fellows}, Einstein\footnote{http://cxc.harvard.edu/fellows/fellowslist.html}, and Sagan\footnote{http://nexsci.caltech.edu/sagan/postdocRecipients.shtml} websites. This time range was chosen to provide a significant pool of recipients while covering a consistent time period of fellowship operations. For each recipient, we aimed to gather their name, gender, and year of PhD.

From the listings on the fellowship websites, we obtained the name and award year for each recipient. On the Einstein fellowship site, short biosketches of each recipient are provided, with third-person pronouns included, which we used to assign binary gender (male or female; we are not able to categorize individuals who might identify as gender nonconforming). On the Sagan site, photos and first or third-person biosketches are provided; we did our best to assign gender based on all the information listed. The Hubble site did not include any information beyond the name and institutions of the recipients.

In order to gather the gender and year of PhD for all recipients where that information was not listed on the award sites, we searched online for personal web pages of the recipients, or other biographical information listed for their graduate institutions or other research-related websites. From those websites, we gathered the gender and PhD year. We were able to assemble that information for all recipients.

\section{Publication Data}

For each fellowship recipient, we searched for their publication record in the NASA Astrophysical Data System (ADS). Because of different features and ease of use, we use both the classic ADS interface, as well as the newer “Bumblebee” interface. Our goal was to obtain a record of the publications of each recipient at the time they applied for the fellowship. We entered the recipient’s name in the new ADS interface using their full name inverted (last name, first name), applying the “All refereed articles” filter, and creating a time cutoff at November of the year prior to the award year (e.g. Nov. 2016 for the 2017 recipients). 

We did not typically impose an early-time cutoff. However, in certain cases, the query results delivered a large set of relatively recent papers, along with a set of papers under the same name but separated by many years. In those cases, we interpreted the results as indicating two separate people and, therefore, we excluded the earlier set of papers. 

When a simple name search within an appropriate time frame returned a questionably long or short number of publications, or publications which seemed out of the scholar’s research area, we checked if the recipient had a personal CV posted online, either on a personal website or another professional site.  In these situations, we consulted the recipient’s CV for further information to rerun the query with more specific criteria. When a publication list seemed unreasonably long, we used the recipient’s affiliation information to refine the search.  In some cases, in order to improve or double-check the searches, we reran the query using variations of the name (initials or common variations of the first name) found within the CV, which returned additional publications or removed those with incorrect name variations. If all this was insufficient, we directly examined the publications listed on the CV to identify correct publications within the query.  We found that 70\% of the names required additional search criteria to correctly isolate the publications of the recipient, such as affiliation, full name, or other restrictions.

We attempted to look for any cases where the recipient name may have changed during their scientific career, such as name changes related to marriage.  We found only one clear such case, and used the version of the name before and after the change to assemble the publication list.

The final retrieved publication data contained the total number of refereed publications for the recipient, differentiating between first-authorships and co-authorships. To enable rechecking of these results, we recorded the URL of every personal or affiliated site where relevant information was found, the list of publications in ADS classic format, and the date of data retrieval for each recipient. While this information is public, we do not provide it here in order to maintain some privacy for individuals. We do include the full anonymized data table in the appendix.

\section{Analysis}

The goal of this project is to identify patterns among the numbers of publications that fellowship recipients had at the time of application for the fellowship. We examined these numbers across gender, fellowship year, and fellowship name. These included the years 2014, 2015, 2016, and 2017 for the Einstein, Hubble, and Sagan fellowships.

\begin{figure}[t]
\centering
\includegraphics[width=0.45\textwidth]{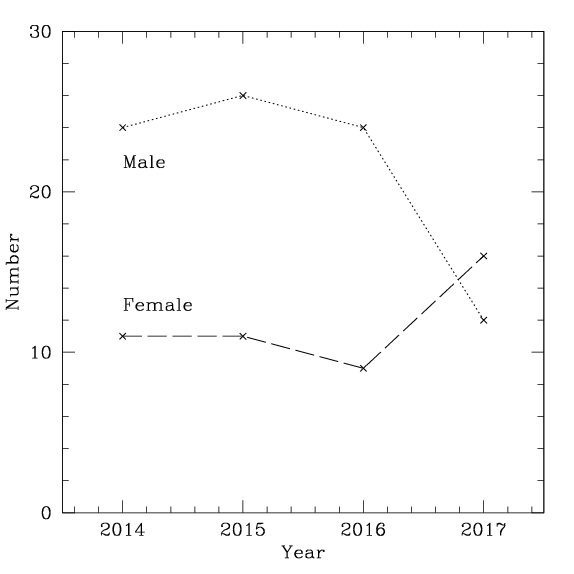}
\caption{Gender breakdown of recipients of all three fellowships across time.}
\label{fig:gen}
\end{figure}

Before examining the publication information, it is worth looking at some of the general demographics. The Hubble fellowship is by far the largest category with 65 recipients, compared to 46 for the Einstein and 22 for the Sagan. All three fellowships skew male, with the Hubble closest to parity at 42\% female, and the Sagan the least, at 27\% female. It is interesting, however, to also examine the gender ratio across time. In Figure \ref{fig:gen}, we see that, while across all fellowships the gender ratio was relatively constant at about 2 to 1 male to female, there seems to have been a significant change starting in 2017.

\begin{figure}[t]
\centering
\includegraphics[width=0.45\textwidth]{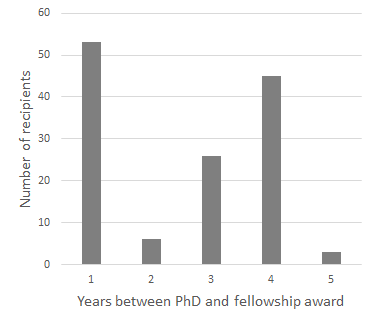}
\caption{Gap in time between receiving a PhD and receiving the fellowship.}
\label{fig:delay}
\end{figure}

Overall, about 40\% of the recipients received the fellowship right out of their PhD. About 20\% more received it after 2 years, and another 30\% after 3 years, as shown in Figure \ref{fig:delay}.  Since we record only whole units of years, a few recipients are listed as receiving the fellowship after more than 3 years, but that only represents recipients who graduated outside of the typical academic cycle. We also examined how many years after their PhD the recipients received the fellowship, using 2 years as a dividing line.

For the Hubble, similar numbers of recipients received the fellowship more than 2 years after their PhD (48\%) compared to less than 2 years after their PhD (52\%). However, for the Einstein, 67\% of recipients received the fellowship less than 2 years after their PhD, and for the Sagan the fraction is 82\%. Those trends of time elapsed between receiving a PhD and fellowship are roughly similar for both genders for the Einstein and Sagan. However, for the Hubble fellowship, there is a significant gender difference. While similar numbers of recipients overall were more than 2 years out from a PhD, compared to those less than 2 years out, about 2/3 of the female recipients received the award more than 2 years after PhD while, for male recipients, 2/3 received the award less than 2 years after PhD.

When it comes to publications, there are some fairly consistent patterns, see Table \ref{tab:pubs}. The typical recipient of these fellowships has between 10 and 20 total publications. For the Einstein, the median number of publications is 11.5, while for the Hubble and Sagan the medians are 17 and 15.5, respectively. For first author publications, Hubble and Einstein recipients typically had 6 papers, while Sagan recipients typically had 4 papers.  For all awards, the mean number of publications is greater than the median, due to a small number of recipients with a very large number of publications.  

\begin{table*}
\caption{Number of publications for fellowship recipients.}
\label{tab:pubs}
\begin{center}
\begin{tabular}{ l c c c|c c c} 
 \hline
 &  \multicolumn{3}{c}{Total Publications} & \multicolumn{3}{c}{First-Author Publications}\\ \hline
 Fellowship         &   Mean    &   Median  & Median Absolute   &   Mean    &   Median  & Median Absolute   \\
 (total recipients) &           &           & Deviation         &           &           & Deviation\\
 \hline
 Hubble (65)    & 19.1  & 17.0  & 7.0 & 6.6 & 6.0 & 2.0 \\
 Einstein (46)  & 14.0  & 11.5  & 4.5 & 5.6 & 6.0 & 1.0 \\
 Sagan (22)     & 19.0  & 15.5  & 4.5 & 4.7 & 4.0 & 1.0 \\
 All (133)      & 17.3  & 14.0  & 6.0 & 5.9 & 6.0 & 2.0 \\
 \hline
\end{tabular}
\end{center}
\end{table*}

\begin{figure}[t]
\centering
\includegraphics[width=0.45\textwidth]{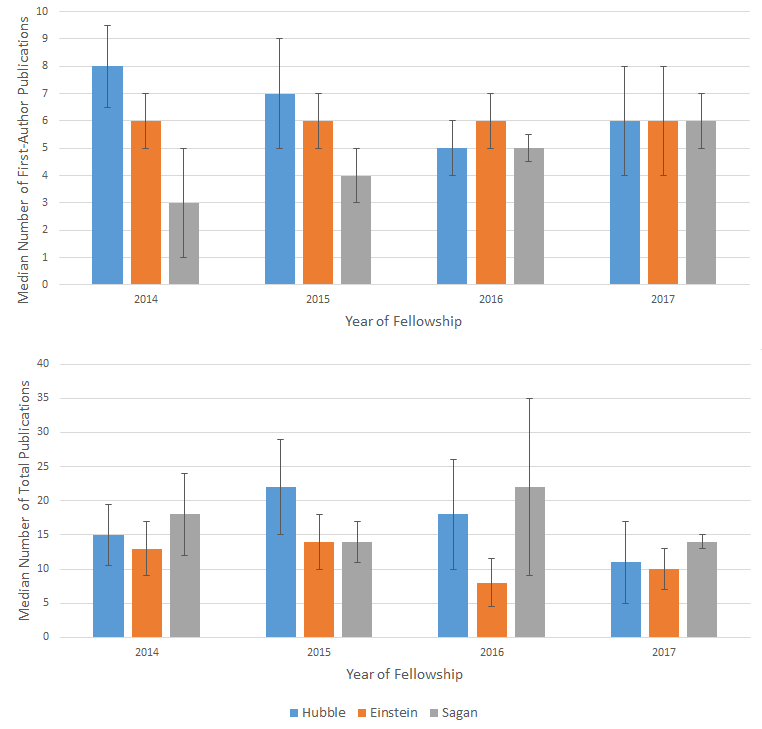}
\caption{Median number of first-author publications (top) and total publications (bottom), broken out by fellowship program.  The error bars indicate the median absolute deviation.}
\label{fig:npub_time}
\end{figure}

\begin{figure}[t]
\centering
\includegraphics[width=0.45\textwidth]{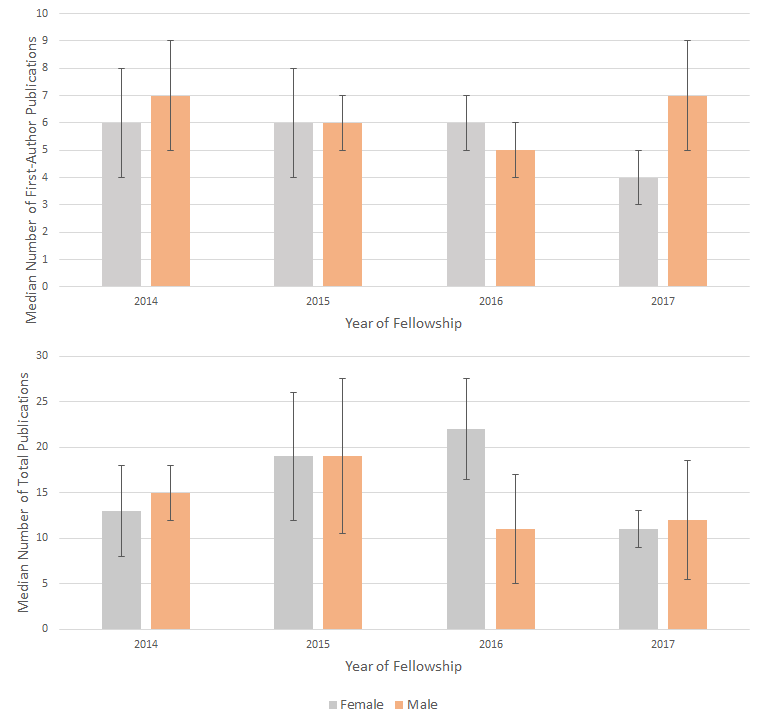}
\caption{Median number of first-author publications (top) and total publications (bottom), broken out by gender of the recipient.  The error bars indicate the median absolute deviation.}
\label{fig:npub_gen}
\end{figure}

We examined how these numbers changed across time in Figure \ref{fig:npub_time}. We found no consistent trend across the four years of the awards considered here in total number of publications.  The median number of first author publications seems to be converging over time, from a range of 3 to 8, to about 5 papers across all three awards.  That convergence is likely partially due to the large fraction of Sagan recipients receiving the fellowship right out of graduate school, compared to the recipients of the other fellowships.  We also examined how the publication numbers break down between male and female recipients.  Those values are shown in Figure \ref{fig:npub_gen}, and we find no consistent difference between those numbers. 

\begin{figure}[t]
\centering
\includegraphics[width=0.45\textwidth]{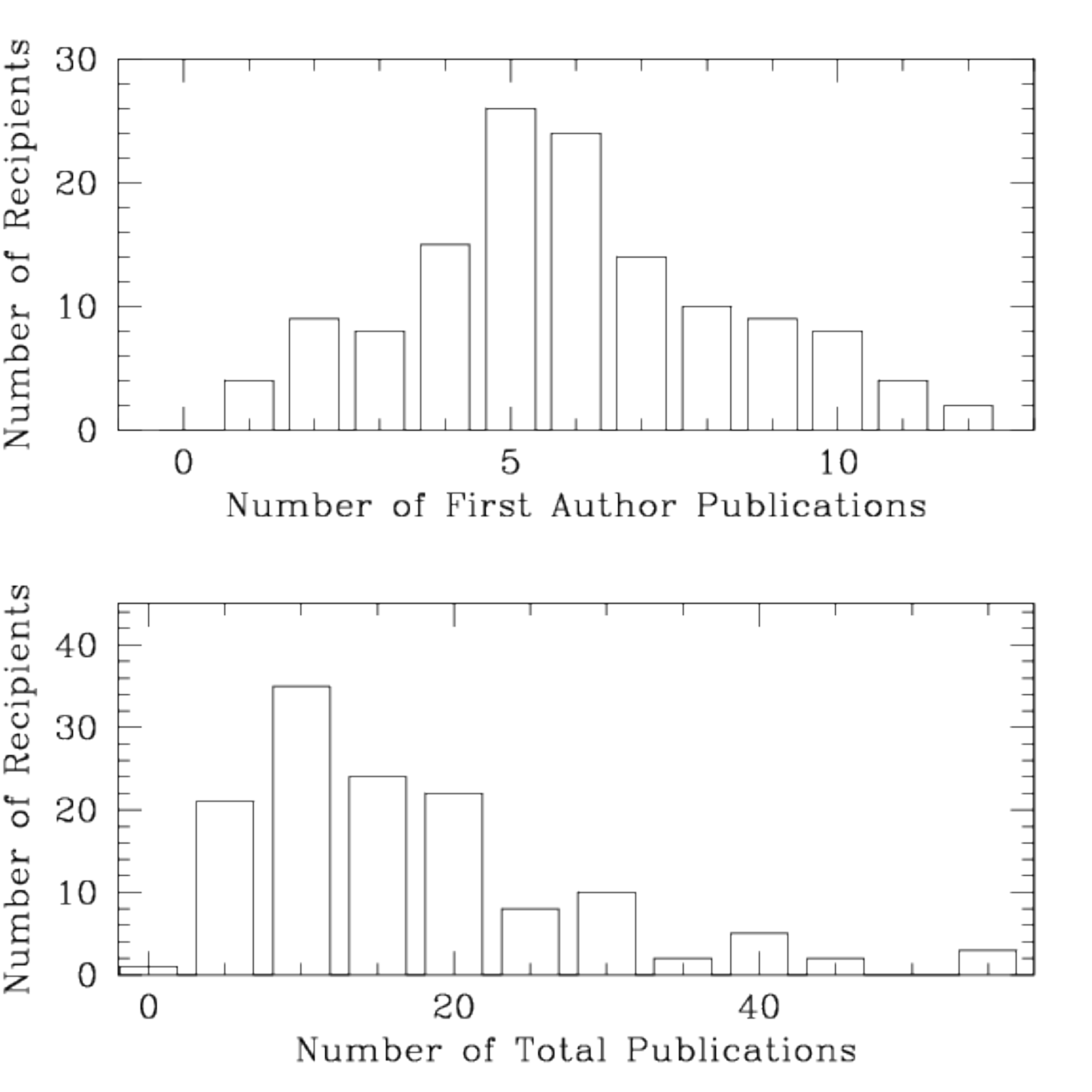}
\caption{Distributions of first author publications and total publications for all recipients.}
\label{fig:bhist}
\end{figure}

To provide a broader overview of the diversity of publication rates, we created histograms of the number of publications for all recipients, as shown in Figure \ref{fig:bhist}, based on the number of publications at the presumed date of application, which we defined as the end of the calendar year prior to receiving the fellowship.  The most noteworthy property of Figure \ref{fig:bhist} is the large spread in the distribution of publications.  While the distributions peak around five and ten first-author and total publications, respectively, the spread is broad, with significant numbers of recipients with three or fewer first author papers upon applying for fellowships.  This demonstrates that while the number of papers may factor into these awards, they are far from deterministic.

One interesting aspect of these distributions is the non-trivial numbers of recipients with very few and very many publications, both for first-author and total publications.  We examined the tails of the publication distributions, looking for any topical or methodological commonalities.  At both the low and high ends, no clear trends emerged, with people involved in theory, observation, instrumentation, and simulation, across all subject areas of astronomy.

\subsection{Immediate Recipients and Later Recipients}

\begin{figure}[t]
\centering
\includegraphics[width=0.45\textwidth]{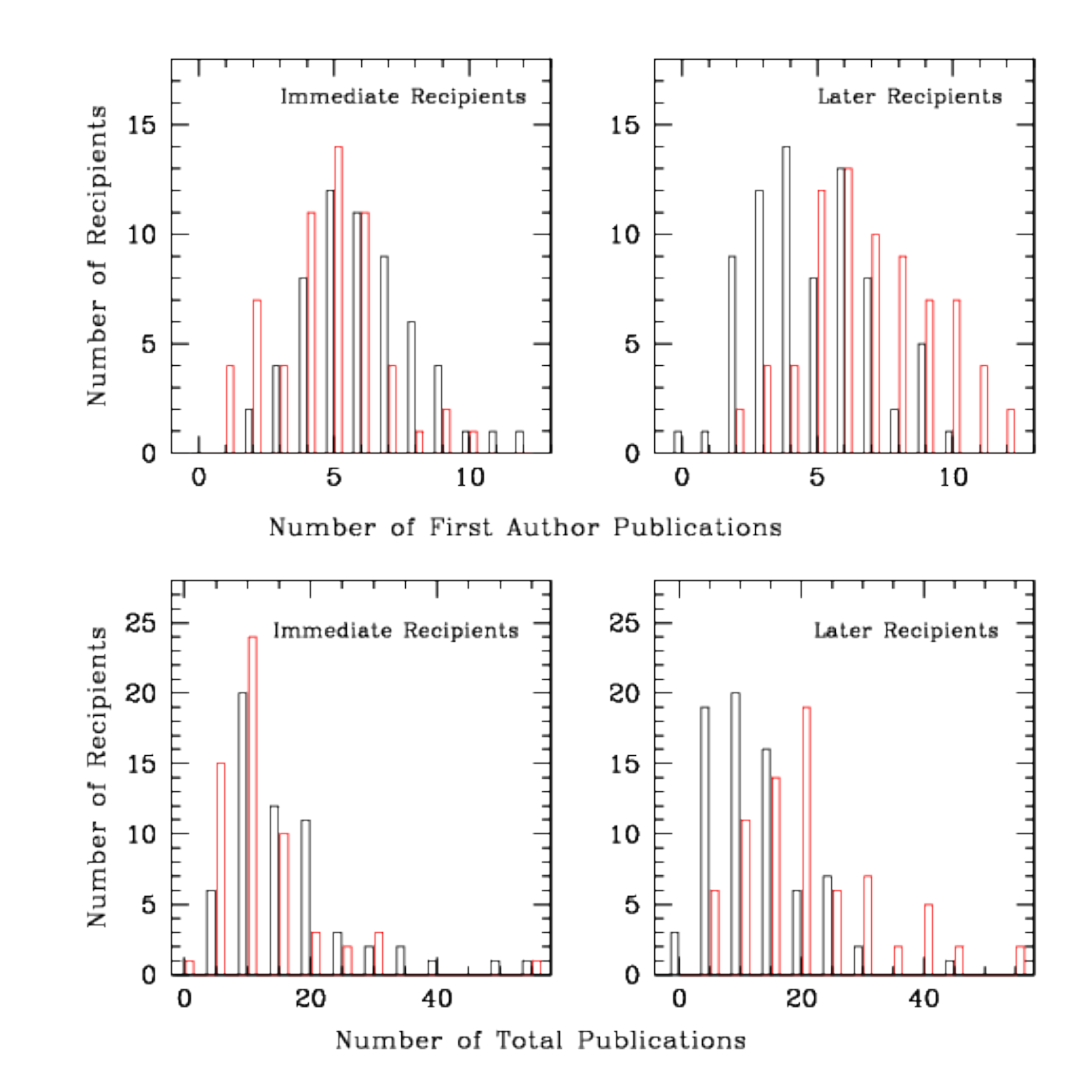}
\caption{Distributions of first author publications and total publications for those receiving the fellowship within a year of PhD (left), and for those who receive the fellowship at least 2 years later (right).  Publications calculated at time of application (red) and at end of the PhD calendar year (black).}
\label{fig:bdhist}
\end{figure}

We can also look at how the publication distribution profile of later-career recipients compares to that of those right out of their PhD, looking at the distributions from Figure \ref{fig:bhist} in more detail.  For that, we plot the distribution of publications separately for those who receive fellowships right out of PhD compared to later recipients, using the 2-year time difference between PhD year and award year to differentiate the populations.\footnote{For example, for someone who received their PhD in 2013 and was awarded the fellowship in 2016, we used November 2015 as the cutoff date for publications, and considered their publications across two years for the per-year metrics.}  Figure \ref{fig:bdhist} shows those distributions, with recipients publications plotted at both the year of PhD and year of application.

There are some interesting patterns seen in Figure \ref{fig:bdhist}.  74 of the applicants, representing 56\% of the total, received the award more than a year after they received their PhD (likely after one postdoctoral position).  We find that those recipients produced an average (median) of 5.9 (4.0) total publications per year in the intervening years, and an average (median) of 1.4 (1.0) first-author publications per year in the intervening years.  Overall, that population increased their publication numbers from $5\pm2$ to $7\pm2$ first author papers, and from $11\pm5$ to $20\pm7$ total papers, using median and median absolute deviations.

Compared to those receiving the fellowship immediately, those who receive it later have more total publications at the point of application: $20\pm7$ for later recipients compared to $10\pm4$ for immediate recipients. However, that is less the case for first author papers: $7\pm2$ for later recipients compared to $5\pm1$ for immediate recipients.

\section{Qualifications and Caveats}

The selection of the three NASA postdoctoral fellowship programs was due primarily for convenience.  It would be useful to include additional postdoctoral fellowship programs to see whether the publication trends seen here are typical.  A future analysis could include the NSF postdoctoral fellowships, the NASA Postdoctoral Program, and other postdoctoral fellowships in astronomy. It is possible that the particular scientific focus of the fellowships considered here excludes certain fields in astronomy with different publication patterns, such as students involved in large collaborations like astroparticle physics.  Also, although the Sagan program, and the Michelson program before that, has awarded fellowships for instrumentation-based work, it is possible that that sector of the astronomy field is underrepresented in the awards.

We gathered demographic and publication information for several years of NASA postdoctoral fellowship recipients, relying on a combination of centralized and decentralized information sources. While the fellowship websites provide lists of recipients, and in some cases list the recipient gender or describe them with gendered pronouns, we sometimes had to infer the gender of the recipients based on photos and names. We also checked personal websites and CVs to confirm those inferences to the best of our ability. For analysis purposes, we chose to categorize gender in a binary fashion (male and female); the study is thus limited in its generalizability to transgender and non-binary communities.
	
The publication information can suffer from several kinds of inconsistency and incompleteness. We only consider dates in units of whole years, and do not attempt to determine publication records on monthly timescales. Therefore, the number of publications for each recipient may not match precisely to the record used in fellowship applications. Furthermore, the ADS system may be missing papers published in less commonly used astronomy journals. In certain cases, the recipients’ names are quite common, and we had to rely on affiliation information to select the papers. While we used personal websites and online CVs posted by many of the recipients to verify the list of papers, we may have missed some publications. 

Finally, this analysis only involves the last four years of the NASA Hubble, Einstein, and Sagan postdoctoral fellowships. While we do not have a reason to believe that these results should differ from other fellowships, there may be differences in how the different programs evaluate applications.

\section{Discussion}

This analysis is intended to provide empirical information about the number of peer-reviewed publications that are typical for recipients of astronomical postdoctoral fellowships. Although publications reflect only one piece of a successful fellowship application, they are generally regarded as a significant part of it, and are one of the few components that can be evaluated in a quantitative way. Of course, publications are only part of an application; successful applications need to have an impactful, cutting edge idea or project that the committee is convinced the applicant can successfully complete in the time allotted by the fellowship.  

We see that the number of papers, especially as displayed in Figure 5, is quite broad, with no evidence for a particular cutoff or minimum number of papers needed, reflecting the variability of a successful record and application.  We hope that this work provides context for discussions between graduate students and their advisors about how to prepare applications for postdoctoral positions

\acknowledgements{
We thank Andrew Fruchter and Paul Green for useful conversations.  This research has made use of NASA’s Astrophysics Data System.  
}

\appendix

\begin{table*}
\caption{Anonymized list of Einstein fellowship recipients with PhD dates and publication information}
\label{tab:ein_awards}
\begin{center}
\begin{tabular}{ c c c c c c c c } 
 & & & &  \multicolumn{4}{c}{Publications} \\
 & & & &  \multicolumn{2}{c}{At PhD year} & \multicolumn{2}{c}{At application date}   \\
 Fellowship &   Award Year  & PhD Year  & Sex   & First Author  & Total & First Author  & Total \\
 \hline
Einstein &	2017 &	2017 &	Female &	5   &	11 &	3   &	7  \\
Einstein &	2017 &	2017 &	Female &	2   &	9  &	2   &	3  \\
Einstein &	2017 &	2014 &	Male   &	3   &	4  &	7   &	10 \\
Einstein &	2017 &	2017 &	Male   &	13  &	17 &	10  &	13 \\
Einstein &	2017 &	2017 &	Male   &	7   &	19 &	5   &	15 \\
Einstein &	2017 &	2017 &	Female &	4   &	4  &	2   &	2  \\
Einstein &	2017 &	2016 &	Male   &	7   &	10 &	7   &	10 \\
Einstein &	2017 &	2017 &	Female &	6   &	11 &	6   &	11 \\
Einstein &	2016 &	2013 &	Male   &	5   &	47 &	6   &	71 \\
Einstein &	2016 &	2014 &	Female &	2   &	3  &	5   &	7  \\
Einstein &	2016 &	2016 &	Male   &	9   &	9  &	5   &	5  \\
Einstein &	2016 &	2016 &	Male   &	7   &	8  &	5   &	5  \\
Einstein &	2016 &	2016 &	Male   &	6   &	9  &	2   &	4  \\
Einstein &	2016 &	2014 &	Male   &	6   &	6  &	6   &	8  \\
Einstein &	2016 &	2016 &	Male   &	8   &	12 &	5   &	10 \\
Einstein &	2016 &	2014 &	Male   &	5   &	5  &	8   &	8  \\
Einstein &	2016 &	2016 &	Female &	5   &	7  &	1   &	3  \\
Einstein &	2016 &	2013 &	Male   &	6   &	24 &	7   &	38 \\
Einstein &	2016 &	2015 &	Male   &	9   &	15 &	9   &	15 \\
Einstein &	2016 &	2013 &	Male   &	4   &	10 &	8   &	21 \\
Einstein &	2015 &	2013 &	Female &	1   &	1  &	7   &	61 \\
Einstein &	2015 &	2011 &	Male   &	4   &	4  &	6   &	12 \\
Einstein &	2015 &	2013 &	Female &	5   &	14 &	6   &	16 \\
Einstein &	2015 &	2015 &	Male   &	10  &	22 &	8   &	15 \\
Einstein &	2015 &	2015 &	Male   &	4   &	17 &	4   &	12 \\
Einstein &	2015 &	2015 &	Male   &	11  &	15 &	9   &	13 \\
Einstein &	2015 &	2015 &	Male   &	6   &	29 &	5   &	10 \\
Einstein &	2015 &	2015 &	Male   &	4   &	18 &	4   &	18 \\
Einstein &	2015 &	2012 &	Male   &	3   &	15 &	5   &	26 \\
Einstein &	2015 &	2015 &	Female &	3   &	10 &	3   &	10 \\
Einstein &	2015 &	2012 &	Male   &	6   &	6  &	6   &	7  \\
Einstein &	2015 &	2012 &	Female &	4   &	15 &	5   &	21 \\
Einstein &	2015 &	2012 &	Male   &	8   &	18 &	11  &	31 \\
Einstein &	2015 &	2013 &	Male   &	5   &	6  &	5   &	6  \\
Einstein &	2014 &	2014 &	Male   &	2   &	22 &	1   &	15 \\
Einstein &	2014 &	2014 &	Female &	9   &	16 &	7   &	10 \\
Einstein &	2014 &	2014 &	Male   &	5   &	14 &	5   &	10 \\
Einstein &	2014 &	2014 &	Female &	7   &	27 &	5   &	18 \\
Einstein &	2014 &	2011 &	Male   &	3   &	7  &	7   &	12 \\
Einstein &	2014 &	2012 &	Male   &	4   &	6  &	8   &	14 \\
Einstein &	2014 &	2011 &	Male   &	2   &	2  &	5   &	5  \\
Einstein &	2014 &	2014 &	Female &	8   &	22 &	6   &	14 \\
Einstein &	2014 &	2011 &	Male   &	6   &	13 &	9   &	20 \\
Einstein &	2014 &	2014 &	Male   &	6   &	9  &	6   &	7  \\
Einstein &	2014 &	2011 &	Male   &	5   &	12 &	7   &	20 \\
Einstein &	2014 &	2011 &	Female &	2   &	2  &	2   &	6  \\
\hline
\end{tabular}
\end{center}
\end{table*}

\begin{table*}
\caption{Anonymized list of Hubble fellowship recipients with PhD dates and publication information}
\label{tab:hub1_awards}
\begin{center}
\begin{tabular}{ c c c c c c c c } 
 & & & &  \multicolumn{4}{c}{Publications} \\
 & & & &  \multicolumn{2}{c}{At PhD year} & \multicolumn{2}{c}{At application date}   \\
 Fellowship &   Award Year  & PhD Year  & Sex   & First Author  & Total & First Author  & Total \\
 \hline
Hubble	 &	2017 &	2014 &	Female &	3   &	28 &	3   &	31 \\
Hubble	 &	2017 &	2017 &	Male   &	5   &	16 &	5   &	11 \\
Hubble	 &	2017 &	2013 &	Female &	7   &	8  &	10  &	11 \\
Hubble	 &	2017 &	2017 &	Female &	3   &	6  &	2   &	5  \\
Hubble	 &	2017 &	2014 &	Female &	7   &	27 &	9   &	41 \\
Hubble	 &	2017 &	2017 &	Female &	5   &	8  &	4   &	4  \\
Hubble	 &	2017 &	2014 &	Female &	4   &	9  &	5   &	12 \\
Hubble	 &	2017 &	2017 &	Male   &	6   &	13 &	6   &	10 \\
Hubble	 &	2017 &	2017 &	Male   &	5   &	10 &	7   &	8  \\
Hubble	 &	2017 &	2013 &	Female &	3   &	8  &	6   &	18 \\
Hubble	 &	2017 &	2015 &	Male   &	6   &	13 &	8   &	16 \\
Hubble	 &	2017 &	2017 &	Female &	4   &	11 &	2   &	9  \\
Hubble	 &	2017 &	2015 &	Male   &	2   &	6  &	5   &	11 \\
Hubble	 &	2017 &	2014 &	Female &	7   &	12 &	10  &	30 \\
Hubble	 &	2017 &	2017 &	Female &	9   &	11 &	4   &	5  \\
Hubble	 &	2017 &	2015 &	Female &	5   &	13 &	6   &	18 \\
Hubble	 &	2017 &	2015 &	Male   &	9   &	28 &	11  &	41 \\
Hubble	 &	2016 &	2013 &	Male   &	10  &	27 &	15  &	41 \\
Hubble	 &	2016 &	2015 &	Male   &	5   &	11 &	5   &	10 \\
Hubble	 &	2016 &	2013 &	Female &	6   &	17 &	6   &	22 \\
Hubble	 &	2016 &	2016 &	Female &	7   &	37 &	6   &	28 \\
Hubble	 &	2016 &	2016 &	Male   &	5   &	5  &	4   &	4  \\
Hubble	 &	2016 &	2016 &	Male   &	4   &	11 &	4   &	9  \\
Hubble	 &	2016 &	2014 &	Male   &	4   &	12 &	5   &	13 \\
Hubble	 &	2016 &	2013 &	Female &	3   &	6  &	6   &	17 \\
Hubble	 &	2016 &	2016 &	Male   &	3   &	22 &	1   &	11 \\
Hubble	 &	2016 &	2013 &	Male   &	5   &	11 &	8   &	18 \\
Hubble	 &	2016 &	2013 &	Female &	3   &	8  &	6   &	21 \\
Hubble	 &	2016 &	2013 &	Female &	5   &	25 &	7   &	35 \\
Hubble	 &	2016 &	2016 &	Male   &	6   &	16 &	4   &	10 \\
Hubble	 &	2016 &	2013 &	Male   &	2   &	7  &	4   &	22 \\
Hubble	 &	2016 &	2014 &	Male   &	4   &	18 &	5   &	34 \\
\hline
\end{tabular}
\end{center}
\end{table*}

\begin{table*}
\caption{Table \ref{tab:hub1_awards} continued}
\label{tab:hub2_awards}
\begin{center}
\begin{tabular}{ c c c c c c c c } 
 & & & &  \multicolumn{4}{c}{Publications} \\
 & & & &  \multicolumn{2}{c}{At PhD year} & \multicolumn{2}{c}{At application date}   \\
 Fellowship &   Award Year  & PhD Year  & Sex   & First Author  & Total & First Author  & Total \\
 \hline
Hubble	 &	2015 &	2012 &	Female &	2   &	23 &	6   &	46 \\
Hubble	 &	2015 &	2012 &	Male   &	2   &	14 &	10  &	41 \\
Hubble	 &	2015 &	2013 &	Male   &	3   &	18 &	5   &	22 \\
Hubble	 &	2015 &	2012 &	Female &	0   &	3  &	2   &	7  \\
Hubble	 &	2015 &	2015 &	Female &	8   &	15 &	5   &	9  \\
Hubble	 &	2015 &	2013 &	Male   &	9   &	23 &	12  &	32 \\
Hubble	 &	2015 &	2013 &	Male   &	6   &	16 &	7   &	20 \\
Hubble	 &	2015 &	2012 &	Male   &	4   &	7  &	6   &	13 \\
Hubble	 &	2015 &	2015 &	Female &	8   &	35 &	6   &	25 \\
Hubble	 &	2015 &	2012 &	Male   &	7   &	13 &	10  &	24 \\
Hubble	 &	2015 &	2013 &	Male   &	9   &	14 &	11  &	22 \\
Hubble	 &	2015 &	2012 &	Male   &	9   &	10 &	11  &	21 \\
Hubble	 &	2015 &	2013 &	Male   &	9   &	20 &	10  &	29 \\
Hubble	 &	2015 &	2015 &	Male   &	4   &	6  &	6   &	12 \\
Hubble	 &	2015 &	2013 &	Male   &	4   &	18 &	7   &	28 \\
Hubble	 &	2015 &	2012 &	Male   &	4   &	7  &	6   &	22 \\
Hubble	 &	2015 &	2012 &	Female &	6   &	24 &	9   &	47 \\
Hubble	 &	2014 &	2014 &	Female &	6   &	18 &	3   &	9  \\
Hubble	 &	2014 &	2014 &	Male   &	5   &	14 &	4   &	10 \\
Hubble	 &	2014 &	2011 &	Female &	4   &	11 &	5   &	19 \\
Hubble	 &	2014 &	2011 &	Male   &	7   &	7  &	9   &	12 \\
Hubble	 &	2014 &	2011 &	Female &	7   &	9  &	9   &	14 \\
Hubble	 &	2014 &	2011 &	Male   &	4   &	16 &	7   &	25 \\
Hubble	 &	2014 &	2011 &	Female &	6   &	10 &	8   &	15 \\
Hubble	 &	2014 &	2011 &	Male   &	6   &	8  &	10  &	17 \\
Hubble	 &	2014 &	2012 &	Male   &	6   &	17 &	7   &	20 \\
Hubble	 &	2014 &	2013 &	Female &	6   &	8  &	6   &	8  \\
Hubble	 &	2014 &	2011 &	Male   &	4   &	8  &	9   &	15 \\
Hubble	 &	2014 &	2011 &	Female &	6   &	9  &	8   &	11 \\
Hubble	 &	2014 &	2014 &	Male   &	6   &	26 &	4   &	15 \\
Hubble	 &	2014 &	2011 &	Male   &	7   &	10 &	8   &	23 \\
Hubble	 &	2014 &	2014 &	Male   &	7   &	21 &	5   &	8  \\
Hubble	 &	2014 &	2011 &	Male   &	7   &	14 &	9   &	25 \\
\hline
\end{tabular}
\end{center}
\end{table*}

\begin{table*}
\caption{Anonymized list of Sagan fellowship recipients with PhD dates and publication information}
\label{tab:sag_awards}
\begin{center}
\begin{tabular}{ c c c c c c c c } 
 & & & &  \multicolumn{4}{c}{Publications} \\
 & & & &  \multicolumn{2}{c}{At PhD year} & \multicolumn{2}{c}{At application date}   \\
 Fellowship &   Award Year  & PhD Year  & Sex   & First Author  & Total & First Author  & Total \\
 \hline
Sagan	 &	2017 &	2015 &	Female &	3   &	11 &	4   &	14 \\
Sagan	 &	2017 &	2017 &	Male   &	5   &	9  &	6   &	13 \\
Sagan	 &	2017 &	2017 &	Male   &	8   &	52 &	7   &	27 \\
Sagan	 &	2016 &	2014 &	Female &	3   &	19 &	4   &	32 \\
Sagan	 &	2016 &	2015 &	Male   &	5   &	6  &	5   &	6  \\
Sagan	 &	2016 &	2016 &	Male   &	4   &	9  &	2   &	6  \\
Sagan	 &	2016 &	2016 &	Male   &	7   &	90 &	6   &	76 \\
Sagan	 &	2016 &	2016 &	Male   &	7   &	21 &	5   &	16 \\
Sagan	 &	2016 &	2016 &	Female &	6   &	40 &	4   &	28 \\
Sagan	 &	2015 &	2015 &	Female &	6   &	19 &	3   &	11 \\
Sagan	 &	2015 &	2015 &	Male   &	4   &	14 &	2   &	6  \\
Sagan	 &	2015 &	2015 &	Male   &	7   &	19 &	4   &	12 \\
Sagan	 &	2015 &	2013 &	Male   &	6   &	16 &	8   &	20 \\
Sagan	 &	2015 &	2012 &	Male   &	3   &	6  &	5   &	15 \\
Sagan	 &	2015 &	2012 &	Female &	4   &	9  &	4   &	17 \\
Sagan	 &	2014 &	2012 &	Female &	2   &	7  &	3   &	13 \\
Sagan	 &	2014 &	2013 &	Male   &	5   &	28 &	5   &	28 \\
Sagan	 &	2014 &	2012 &	Male   &	8   &	14 &	10  &	18 \\
Sagan	 &	2014 &	2014 &	Male   &	3   &	9  &	1   &	5  \\
Sagan	 &	2014 &	2011 &	Male   &	2   &	3  &	3   &	25 \\
Sagan	 &	2014 &	2014 &	Male   &	8   &	26 &	6   &	19 \\
Sagan	 &	2014 &	2012 &	Male   &	3   &	10 &	3   &	12 \\
\hline
\end{tabular}
\end{center}
\end{table*}

\bibliography{main}

\end{document}